# Enhancing Clustering Algorithm to Plan Efficient Mobile Network

Lamiaa Fattouh Ibrahim*
Department of Computer Sciences and
Information System, Institute of Statistical
Studies and Research, Cairo University Giza,
Egypt

Manal El Harby
College of Education,
UMM AL-QURA University,
Macca, Saudi Arabia

*Corresponding Author: Department of Information Technology,
Faculty of Computer and Information Technology,
King Abdulaziz University
B.P. 42808 Zip Code 21551- Girl Section, Jeddah, Saudi Arabia

## ABSTRACT
With the rapid development in mobile network effective network planning tool is needed to satisfy the need of customers. However, deciding upon the optimum placement for the base stations (BS) to achieve best services while reducing the cost is a complex task requiring vast computational resource. This paper addresses antenna placement problem or the cell planning problem, involves locating and configuring infrastructure for mobile networks. The Cluster Partitioning Around Medoids (PAM) original algorithm has been modified and a new algorithm M-PAM (Modified-Partitioning Around Medoids) has been proposed by the authors in a recent work. In the present paper, the M-PAM algorithm is modified and a new algorithm CWN-PAM (Clustering with Weighted Node-Partitioning Around Medoids) has been proposed to satisfy the requirements and constraints. Implementation of this algorithm to a real case study is presented. Results demonstrate the effectiveness and flexibility of the modifying algorithm in tackling the important problem of mobile network planning.

## 1. INTRODUCTION
The design objective of early mobile radio systems was to achieve a large coverage area by using a single, high powered transmitter with an antenna mounted on a tall tower. While this approach achieved very good coverage, it also meant that it was impossible to reuse those same frequencies throughout the system, since any attempts to achieve frequency reuse would result in interference. Faced with the fact that government regulatory agencies could not make spectrum allocations in proportion to the increasing demand for mobile services, it became imperative to restructure the radio telephone system to achieve high capacity with limited radio spectrum while at the same time covering very large areas [1].

Cellular telephony is designed to provide communications between two moving units, called mobile stations (MSs), or between one mobile unit and one stationary unit, often called a land unit [2]. A service provider must be able to locate and track a caller, assign a channel to the call, and transfer the channel from base station to base station as the caller moves out of range. Each cellular service area is divided into regions called cells. Each cell contains an antenna and is controlled by a solar or AC power network station, called the base station (BS). Each base station, in turn, is controlled by a switching office, called a mobile switching center (MSC). The MSC coordinates communication between all the base stations and telephone central office.

Cell planning is challenging due to inherent complexity, which stems from requirements concerning radio modeling and optimization. Manual human design alone is of limited use in creating highly optimized networks, and it is imperative that intelligent computerized technology [3], Tuba Search TS [4], [5] Genetic Algorithm GA [6] and clustering algorithm [7]-[9] have been successfully deployed in mobile network designs. Clustering analysis is a sub-field in data mining that specializes in techniques for finding similar groups in large database [10]. Its objective is to assign to the same cluster data that are more close (similar) to each other than they are to data of different clusters. The application of clustering in spatial databases presents important characteristics. Spatial databases usually contain very large numbers of points. Thus, algorithms for clustering in spatial databases do not assume that the entire database can be held in main memory. Therefore, additionally to the good quality of clustering, their scalability to the size of the database is of the same importance [11]. In spatial databases, objects are characterized by their position in the Euclidean space and, naturally, dissimilarity between two objects is defined by their Euclidean distance [12]. Clustering techniques has been successfully deployed in wire [13],[14],[15] and in Wireless Local Loop [16] network planning.

This paper introduces the spatial clustering to solve the Mobile Networking Planning problem. This paper is an extension version of papers [17] and [18]. Section 2 discusses main phases used in radio network planning. In sections 3 The Cluster Partitioning Around Medoids (PAM) and M-PAM are reviewed. In section 4, the proposed CWN-PAM algorithm is fully described. A case study is presented in section 5. Section 6 compare between proposed method and other methods. The paper conclusion and Future Work is presented in section 7.

## 2. MAIN PHASES USED IN RADIO NETWORK PLANNING

The radio network planning process can be divided into different phases [19]. At the beginning is the Preplanning phase. In this phase, the basic general properties of the future network are investigated, for example, what kind of mobile services will be offered by the network, what kind of requirements the different services impose on the network, the basic network configuration parameters and so on. The second phase is the main phase. A site survey is done about the to-be-covered area, and the possible sites to set up the base stations are investigated. All the data related to the geographical properties and the estimated traffic volumes at different points of the area will be incorporated into a digital map, which consists of different pixels, each of which records all the information about this point. Based on the propagation model, the link budget is calculated, which will help to define the cell range and coverage threshold. There are some important parameters which greatly influence the link budget, for example, the sensitivity and antenna gain of the mobile equipment and the base station, the cable loss, the fade margin etc. Based on the digital map and the link budget, computer simulations will evaluate the different possibilities to build up the radio network part by using some optimization algorithms. The goal is to achieve as much coverage as possible with the optimal capacity, while reducing the costs also as much as possible. The coverage and the capacity planning are of essential importance in the whole radio network planning. The coverage planning determines the service range, and the capacity planning determines the number of to-be-used base stations and their respective capacities.

In the third phase, constant adjustment will be made to improve the network planning. Through driving tests the simulated results will be examined and refined until the best compromise between all of the facts is achieved. Then the final radio plan is ready to be deployed in the area to be covered and served.

## 3. THE CLUSTER PARTITIONING AROUND MEDOIDS (PAM) AND M-PAM ALGORITHMS

The PAM (Partioning Around Medoids) algorithm, also called the K-medoids algorithm, represents a cluster by a medoid [20]. Initially, the number of desired clusters is input and a random set of k items is taken to be the set of medoids. Then at each step, all items from the input dataset that are not currently medoids are examined one by one to see if they should be medoids. That is, the algorithm determines whether there is an item that should replace one of the existing medoids. By looking at all pairs of medoids, non-medoids objects, the algorithm chooses the pair that improves the overall quality of the clustering the best and exchanges them. Quality here is measured by the sum of all distances from a non-medoid object to the medoid for the cluster it is in. A item is assigned to the cluster represented by the medoid to which it is closest (minimum distance or direct Euclidean distance between the customers and the center of the cluster they belong to).

The PAM algorithm [20] is shown in Figure 1. By assuming that $K_i$ is the cluster represented by medoid $t_i$. Suppose $t_i$ is a current medoid and it is wished to determine whether it should be exchanged with a non-medoid $t_h$. Do this swap only if the overall impact to the cost (sum of the distances to cluster medoids) represents an improvement.

The total impact to quality by a medoid change is given by TC:

$$TC = \sum_{h=1}^{k} \sum_{n_i \in C_i} dis(n_h, n_i) \quad (1)$$

PAM needs to specify number of clusters (k) before starting to search for the best locations of base stations. The M-PAM algorithm uses the radio network planning algorithms to determine the initial k. M-PAM uses also equation 1 to calculate TC.

## 4. THE CWN-PAM ALGORITHM

In a given area, contains number of subscribers, the number of base stations and their boundaries are needed to be determined to satisfy good grade of service with minimum cost.

---

**Algorithm PAM**
**Input**:
D = {$t_1$, $t_2$, $t_3$, ………., $t_n$} // set of elements
 A // adjacency matrix showing distance between
   elements.
k // Number of desired clusters.
**Output**:
K // set of clusters.
**PAM Algorithm:**
 Arbitrarily select k medoids from D;
 repeat
  For each $t_h$ not a medoid do
   For each medoid $t_i$ do
    Compute square error function $TC_{ih}$;
    Find i, h where $TC_{ih}$ is the smallest;
     If $TC_{ih}$ < Current TC then
       Replace medoid $t_i$ with $t_h$;
  Until $TC_{ih}$ >= Current TC;
 For each $t_i \in$ D do
  Assign $t_i$ to $K_j$ where $dis(t_i, t_j)$ is the smallest over all medoids;

---

**Figure 1: PAM algorithm**

**The problem statement:-**
- A set **P** data points {$p_1$, $p_2$… $p_n$} in 2-D map, subscribers loads and communication constraints .
- **Objective** : Partition the city into ***k*** clusters {$C_1$,$C_2$, .., $C_k$} that satisfy clustering constraints, such that the cost function is minimized with high grade of services.
- **Input**: Set of ***n*** objects (map), set of streets associate with their load.

- **Output**: *k* clusters, Base Station locations, boundaries of each cluster.

The proposed algorithm contains three phases. The following sections describe these three phases.

### 4.1 Phase I : Pre-Planning
This phase is divided into two steps. Step 1, convert map from raster form to digital form. Step 2, determine the initial number of clusters.

#### 4.1.1 *Map and their Data Entries*
The maps used for planning are scanned images obtained by the user. It needs some preprocessing operations before it is used as a digital map. The streets and intersection nodes on the raster maps, the beginning and ending of each street are transformed into data nodes, defined by their coordinates. The streets themselves are transformed into links between data nodes. The subscriber's loads are considered to be the weights for each node. Figure 2 show map transformation. For each intersection node and street the user can right click to input the characteristics of intersection node (no, name, capacity) or street (street number, street name, street load).

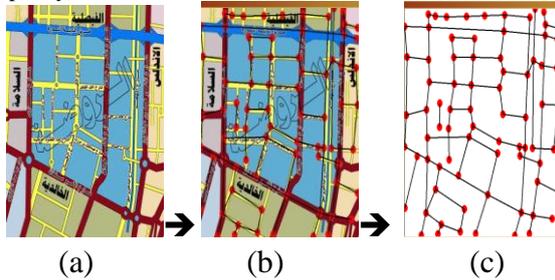

(a)         (b)         (c)

**Figure 2: map transforming**
   (a) raster map
   (b) draw squares and streets on raster map
   (c) map after transforming to digital map

All above data is saved in database. Visual basic dot Net as work environment with oracle is used to create tables and data input.

#### 4.1.2 *Determine Initial Number of Clusters*
In cell planning, planned area is divided to number of cells, each cell served by BS which guarantees the quality of service for all subscribers. In this paper, GSM technology of radio network planning is used to calculate number of cell need by coverage planning and calculate number of cell need by capacity planning for planned area.

- Number of cell needed by coverage planning = Total area / area of the cell
- Number of cell needed by capacity planning= Total number of subscribers / Total subscribers per cell
- Initial number of clusters K = the maximum of the two values.

### 4.2 Phase II : Main-Planning stage
In this phase, the goal is to split the entire database into clusters.

#### 4.2.1 *Partition Database*
After initial k is known, CWN-PAM algorithms is used to determine the optimal location of base station and its boundary of the served area for each cluster. The CWN-PAM algorithm is based mainly on the idea of the Modified Partitioning Around Medoids (M-PAM). The cost function is modified in the M-PAM.

#### 4.2.2 *Modified Cost Function to Handle Node Load*
In database contains n points {$n_1, n_2… n_n$}, Where $n_h$ is the medoid (the real data point that satisfies minimum cost) of cluster $C_h$, $n_i$ denote to non-medoid points and k is the number of cluster. The Direct Euclidean distance from a point $n_h$ to $n_i$ is the dis($n_h$, $n_i$). The cost function in M-PAM (equation 1) is modified.The TC is modified to WTC where:

$$WTC= \sum_{h=1}^{k} \sum_{n_i \in C_h} L_{hi} dis(n_h, n_i) \qquad (2)$$

$L_{hi}$ is the subscriber load cost of this distance. According to equation (2), medoids, location of the base stations, move toward the heavy loaded (weighted) nodes. . Figure 3 shown map of a city and number in map represent the subscriber loads. Block 1 and block 3 has homogeneous distribution loads therefore the medoids are not changed when applying M-PAM and CWN-PAM algorithms. Block 2 and 4 has non-homogeneous distribution load therefore medoids are changed locations when applying M-PAM and CWN-PAM algorithms (medoids move toward the heavy loaded nodes.

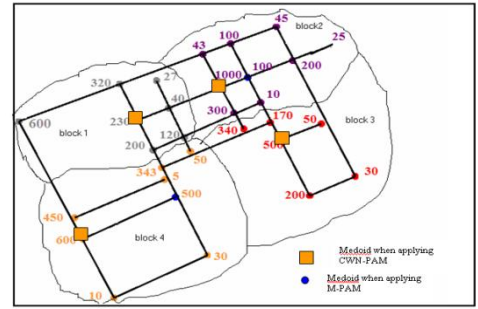

**Figure 3 Map after apply M-PAM and CWN-PAM algorithms**

### 4.3 Phase III : Adjustment Stage
The base station satisfy the constraints if and only if (k<=1) for each cluster. If any of clusters are not satisfy the requirements, more clusters is added and redistribute nodes to close base station.
For each cluster, Coverage and capacity plan are applied to calculate the number of needed base station BS. If any cluster needs more than one base station, one of the following methods is applied. Method I: number of clusters is increased on the whole data (area), and then goes to the Phase II. Method II: number of clusters, on just the cluster that had a problem on its mobile constraints, is increased. Figure 4 describes the new algorithm CWN-PAM.

## 5. CASE STUDY
By applying CWN-PAM algorithm method I and method II to different datasets Table 1 and different cell range the results shown in figure 5 – figure 7 are obtained.

```
Algorithm CWN-PAM
Input:
 D = {t1, t2, t3, ........., tn} // set of elements
 A    // adjacency matrix showing distance between elements.
 Surface of area to be plan
Output:
 A partition of the d objects into k clusters and clusters' centre, m1, m2...... mk
 Cost of planning
CWN-PAM Algorithm:
  Initial K  (Number of desired clusters)
       Number-for-cells1 = Call  capacity algorithm
       Coverage area = Call  coverage algorithm
  Number- of- cells2= Surface of area to be plan / caverage area
  K = max(number-for-cells1 , number-for-cells2)
  Arbitrarily select K medoids from D;
  label 1    repeat
          For each th not a medoid do
           For each medoid ti do
                Compute  error function WTCih;
                Find i, h where WTCih is the smallest;
                If WTCih < current WTCih then
                    Replace medoid ti with th;
           Until WTCih >= current WTCih;
           For each ti ∈ D do
             Assign ti to Kj where dis(ti, tj) is the smallest over all medoids
             For {I = 1 to k }  /* k = number of cluster
             Call  capacity algorithm /* number-for-cells1
             Call  caverage algorithm /* number-for-cells2
 If(number-for-cells1  > 1 or number-for-cells2 >1 ) Then
 If applying method I then   K=K+1  go to label 1
 If applying method II then
    Divide the cluster had a problem into two cluster and
    Apply CWN-PAM algorithm to the divided cluster
```

**Figure 4 CWN-PAM Algorithm**

**Table 1. Data base entries for comparison**

| Data set / Number of Nodes | Coverage Area [$m^2$] | Number of Subscribers |
|---|---|---|
| Data set I = 50 | 230850 | 3139 |
| Data set II = 70 | 335478 | 3500 |
| Data set III = 101 | 337800 | 4000 |
| Data set IV = 150 | 345663 | 4488 |
| Data set V = 300 | 394284 | 10159 |

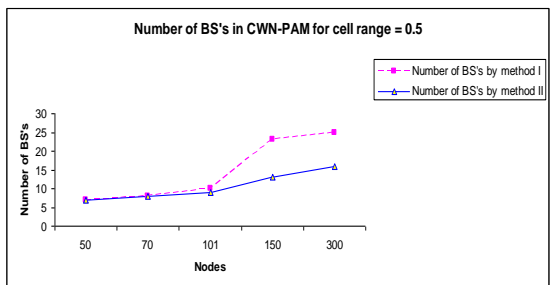

**Figure 5: Number of BS's when applying CWN-PAM (method I Vs. method II) for cell=0.5**

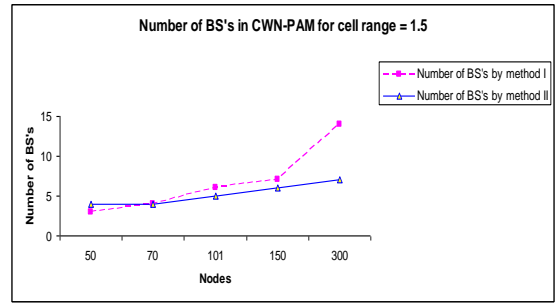

**Figure 6: Number of BS's when applying CWN-PAM (method I Vs. method II) for cell=1.5**

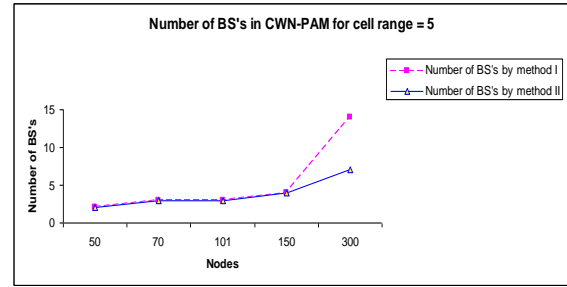

**Figure 7: Number of BS's when applying CWN-PAM (method I Vs. method II) for cell=5**

The experiments show that method II does not have any better solution with small datasets (like dataset I and dataset II) or in the homogeneous density maps. But when large number of subscribers is deployed in different area with different densities, it produces a better solution and cost minimization. Because the clustering algorithm is applied only on the clusters that had a problem on its mobile requirements whether coverage or capacity or even both instead of clustering a whole planned area by adding more clusters and redistribute nodes by using clustering algorithms which sometimes repeat the problem that depending on the redistributing process.

To compare the different densities, two maps with (30,101) nodes are considered. For each map the load of their nodes is changed. The data base is shown in table 2.

**Table 2. Data base entries used for comparison (same maps with different densities)**

| Data set ID | Number of Nodes | Coverage Area | Number of Subscribers |
|---|---|---|---|
| Data set I | 30 | 327600 [$m^2$] | 3000 |
| Data set II | 30 | 327600 [$m^2$] | 7158 |
| Data set III | 101 | 337800 [$m^2$] | 4000 |
| Data set IV | 101 | 337800 [$m^2$] | 18595 |

Figure 8 to 11 show the results of applying CWN-PAM algorithm method I and method II to numbers of nodes 30 and 101 but when increase density (numbers of subscribers).

Dataset I with 30 nodes and 3000 subscribers and dataset III with 101 nodes and 4000 subscribers, since the number of subscribers is small; the number of base stations had different values with different cells range. Dataset II with 30 nodes and 7158 subscribers and dataset IV with 101 nodes and 18595 subscribers, since the number of subscribers is big, therefore the problem in these maps is in their capacity. The number of base stations had not based effected with different cells ranges. But also method II obtain the minimum numbers of Base station.

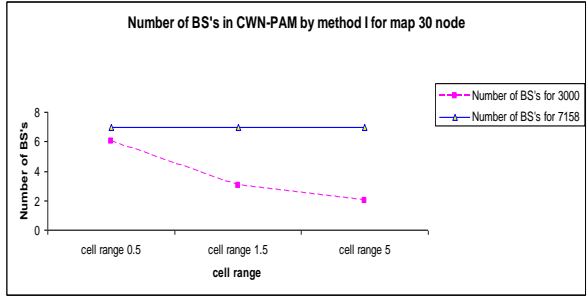

**Figure 8 Number of BS's in CWN-PAM by method I for 30 nodes**

The comparison between applying CWN-PAM (method I and II) and M-PAM algorithms are shown in Figure 12, Figure 13 and Figure 14. The experiments show that CWN-PAM does not have any better solution with small datasets (like dataset I) or in the homogeneous density maps (like dataset III). But for datasets with number of subscribers are deployed in different area with different densities (like dataset V), it produces a cost minimization and grade of service (signal strong) is high beside the heavy loads.

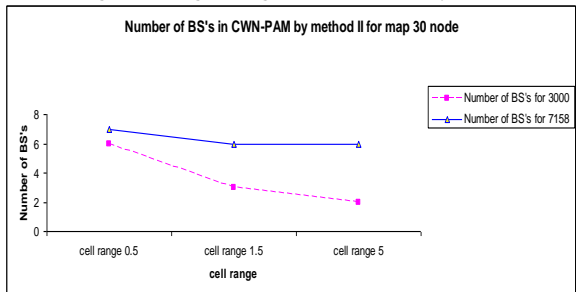

**Figure 9: Number of BS's in CWN-PAM by method II for 30 nodes**

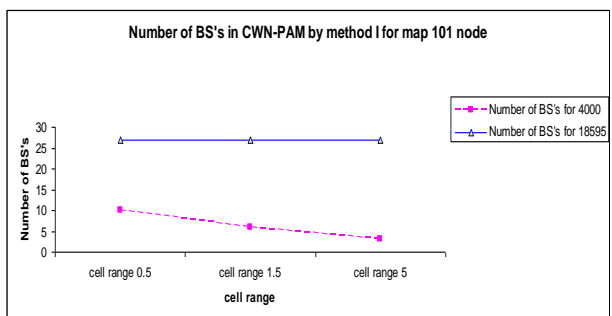

**Figure 10: Number of BS's in CWN-PAM by method I for 101 nodes**

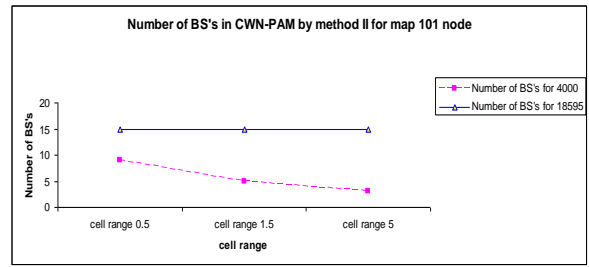

**Figure 11: Number of BS's in CWN-PAM by method II for 101 nodes**

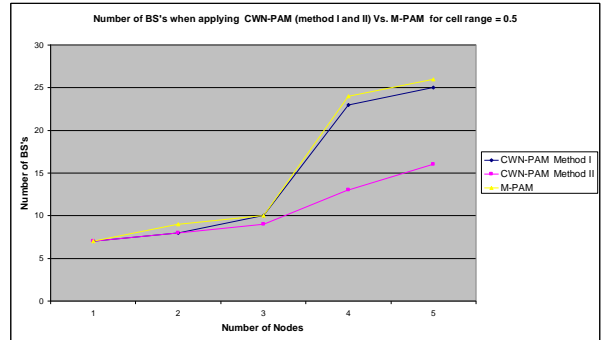

**Figure 12: Comparison between Number of BS's when applying CWN-PAM (method I and II) Vs. M-PAM for cell range = 0.5**

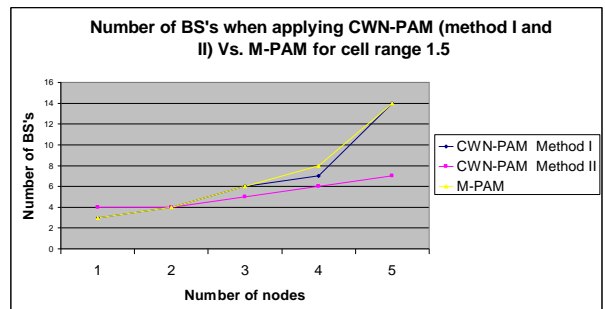

**Figure 13: Comparison between Number of BS's when applying CWN-PAM (method I and II) Vs. M-PAM for cell range = 1.5**

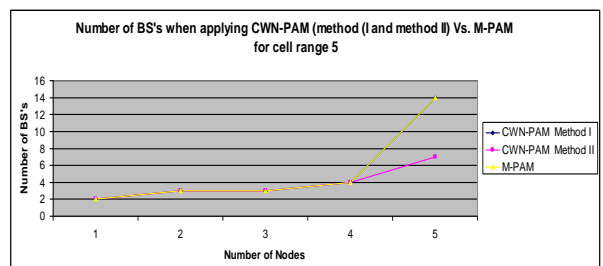

**Figure 14: Comparison between Number of BS's when applying CWN-PAM (method I and II) Vs. M-PAM for cell range = 5**

## 6. COMPARISON BETWEEN PROPOSED METHOD AND OTHER METHODS

Table 3 described the different comparison between the proposed method and other methods using in mobile network planning. There are two methods that are frequently used here: Tabu search [4], [5] and Genetic Algorithms [6].

Table 3 shows the comparison between relative works. Tabu Search and Genetic algorithm, needed a huge numbers of estimated input parameters which can be affected to the results.

## 7. CONCLUSION AND FUTURE WORK

In this paper the proposed algorithm CWN-PAM which use modifies clustering technique M-PAM to the mobile network planning problem has been presented. This algorithm is a medoid clustering algorithm using weighted Direct Euclidean distance that satisfy the network constraints and where the weights used are the subscriber loads. Due to the modification in cost function, the location of the base stations (medoid), move toward the heavy loaded (weighted) nodes and increase there grade of service. Experimental results and analysis indicate that the CWN-PAM algorithm when applying method two is effective, and leads to minimum costs (minimum number of base stations) in network have a large number of intersection and streets where number of subscribers are deployed with different densities. It is expected that by applying this system to a number of areas belonging to different countries with different sizes, one can verify its capabilities more universally.

**Table 3. Relative Works**

| Algorithm | Input Parameters | Results | Location of BS'S | Constraints | Type Of Distance | Complexity |
|---|---|---|---|---|---|---|
| *Genetic Algorithm* | - Data points<br>- Population<br>- Initial probability<br>- Mutation probability<br>- Crossover probability<br>- Number of iteration<br>- Selection pressure | # of clusters | Optimal placement | Fitness Function | Euclidean distance | $O(NK \ln K)$ |
| *Tabu Search* | - Data points<br>- Init probability<br>- Generation probability<br>- Recency factor<br>- Frequency factor<br>- Number of iteration<br>- Number of neighbors | # of clusters | Optimal placement | Tabu List | Euclidean distance | $O(NK \log K)$ |
| *PAM* | - Data points<br>- K | - Clusters medoid<br>- # of clusters | Mediods | $TC_{ih} = \sum C_{ih}$ | Euclidean distance | $O(K(n-K)^2)$ |
| *M-PAM* | - Data points | - Clusters medoid<br>- # of clusters | Mediods | $TC_{ih} = \sum C_{ih}$ | Euclidean distance | $O(K(n-K)^2)$ |
| *CWN-PAM* | -Data points | - Clusters medoid<br>- # of clusters | Mediods | $WTC_{ih} = \sum LC_{ih}$ | Weighted Euclidean distance | $O(K(n-K)^2)$ |